\documentclass[runningheads]{llncs}
\usepackage{graphicx}
\usepackage{float}
\usepackage{marvosym}
\usepackage{xurl}
\begin{document}
\title{Evaluating the Predictive Value of Preoperative MRI for Erectile Dysfunction Following Radical Prostatectomy}
\titlerunning{Predicting Erectile Dysfunction with Preoperative MRI}
\author{Gideon N. L. Rouwendaal\inst{1,2}\textsuperscript{(\Letter)}, Daniël Boeke\inst{2,4,5}, Inge L. Cox\inst{3}, Henk G. van der Poel\inst{3}, Margriet C. van Dijk-de Haan\inst{2}, Regina G. H. Beets-Tan\inst{2,5}, Thierry N. Boellaard\inst{2}\textsuperscript{*}, and Wilson Silva\inst{2,4}\textsuperscript{*}}

\authorrunning{G. Rouwendaal et al.}
\institute{University of Amsterdam, Amsterdam, The Netherlands \email{gideon.rouwendaal@student.uva.nl}  \and Department of Radiology, Antoni van Leeuwenhoek-Netherlands Cancer Institute, Amsterdam, The Netherlands \and Department of Urology, Antoni van Leeuwenhoek-Netherlands Cancer Institute, Amsterdam, The Netherlands \and AI Technology for Life, Department of Information and Computing Sciences, Department of Biology, Utrecht University, Utrecht, The Netherlands \and GROW School for Oncology and Developmental Biology, Maastricht University
Medical Center, Maastricht, The Netherlands}
\maketitle              % typeset the header of the contribution

{\renewcommand{\thefootnote}{\fnsymbol{footnote}}%
\footnotetext[1]{These authors contributed equally as senior authors.}}
\begin{abstract}
Accurate preoperative prediction of erectile dysfunction (ED) is important for counseling patients undergoing radical prostatectomy. While clinical features are established predictors, the added value of preoperative MRI remains underexplored. We investigate whether MRI provides additional predictive value for ED at 12 months post-surgery, evaluating four modeling strategies: (1) a clinical-only baseline, representing current state-of-the-art; (2) classical models using handcrafted anatomical features derived from MRI; (3) deep learning models trained directly on MRI slices; and (4) multimodal fusion of imaging and clinical inputs. Imaging-based models (maximum AUC 0.569) slightly outperformed handcrafted anatomical approaches (AUC 0.554) but fell short of the clinical baseline (AUC 0.663). Fusion models offered marginal gains (AUC 0.586) but did not exceed clinical-only performance. SHAP analysis confirmed that clinical features contributed most to predictive performance. Saliency maps from the best-performing imaging model suggested a predominant focus on anatomically plausible regions, such as the prostate and neurovascular bundles.
While MRI-based models did not improve predictive performance over clinical features, our findings suggest that they try to capture patterns in relevant anatomical structures and may complement clinical predictors in future multimodal approaches.

\end{abstract}

\section{Introduction}

Prostate cancer is one of the most frequently diagnosed cancers in men worldwide, ranking as the second most common cancer among men and the fourth most prevalent overall in 2022 \cite{prostatecancer}. For patients with intermediate- to high-risk localized prostate cancer, Robot-Assisted Laparoscopic Prostatectomy (RALP) is a commonly recommended curative treatment \cite{treatments}. While this surgical approach can improve long-term survival and reduce disease progression \cite{wilt2012radical,bill2018radical,serrell2018review,chierigo2022survival}, it can also lead to side effects such as postoperative urinary incontinence and erectile dysfunction (ED) \cite{lane2022functional}. ED is reported in up to 95\% of men immediately following surgery and may persist in 85\% of cases six years after surgery \cite{lane2022functional}.

Predictors of ED currently established in the literature are primarily clinical, including age, comorbidities, preoperative erectile function, and surgical technique \cite{cn2024surgical,predictors,garcia2015predictive,kilminster2012predicting,meuleman2003erectile,alemozaffar2011prediction}. While recent models based solely on preoperative clinical features achieve reasonable performance (AUCs of 0.74–0.80) \cite{agochukwu2022development,saikali2025development}, they leave room for improvement in achieving more personalized preoperative counseling.

Incorporating additional preoperative data, potentially from other modalities such as imaging, may provide complementary information to enhance predictive performance. In particular, MRI could provide anatomical details of the prostate and surrounding structures, including the neurovascular bundles that are essential for erectile function. Prior studies have demonstrated that features in MRI images, such as prostatic fascia thickness and neurovascular distribution, may indeed hold predictive power for postoperative erectile function outcomes \cite{grivas2019value,van2009role,kleinjan2019prediction}. The prostatic fascia is a tissue that protects the neurovascular bundles, critical for erectile function. When the fascia is thin, the neurovascular bundles are situated closer to the prostate, which limits the surgical margin and increases the risk of nerve injury during radical prostatectomy. Current literature supports the belief that a thicker fascia correlates positively with good postoperative erectile function \cite{grivas2019value,van2009role,kleinjan2019prediction}. However, this belief is based on preoperative, postoperative, and intraoperative analysis, rather than preoperative analysis alone. To the best of our knowledge, no prior work has evaluated whether preoperative MRI, analyzed either via anatomical features or in a deep learning (DL) setting, provides independent predictive value for postoperative ED. We therefore aim to systematically assess the predictive utility of preoperative MRI, benchmarked against a clinical baseline. Our key contributions are as follows:

\begin{enumerate}
    \item We evaluate whether preoperative MRI alone holds predictive value for postoperative ED at 12 months, independently of established clinical predictors. This includes both handcrafted anatomical features, namely fascia thickness and volume, and data-driven DL approaches.
    \item We assess whether combining MRI-derived features with clinical predictors improves performance beyond either modality alone, using multimodal fusion strategies.
    \item We assess whether our DL models focus on clinically relevant anatomical regions using attention-based explainability methods to support the plausibility of MRI-based risk modeling.
\end{enumerate}

\section{Methods}
To investigate whether preoperative MRI can enhance personalized ED prediction beyond established clinical predictors, we adopted a stepwise evaluation strategy. We first established a clinical-only baseline, then assessed the predictive value of MRI via handcrafted anatomical features and end-to-end DL. Lastly, we explored whether combining both modalities improves performance. All models aimed to predict whether the patient developed postoperative ED 12 months after surgery, using a binary target variable. To interpret model behavior, Explainable AI (XAI) techniques, including SHAP for the clinical and fusion models, and attention-based saliency maps for the imaging model, were used to identify which features each model relies on.

\begin{figure}[H]
  \centering
  \includegraphics[width=\linewidth]{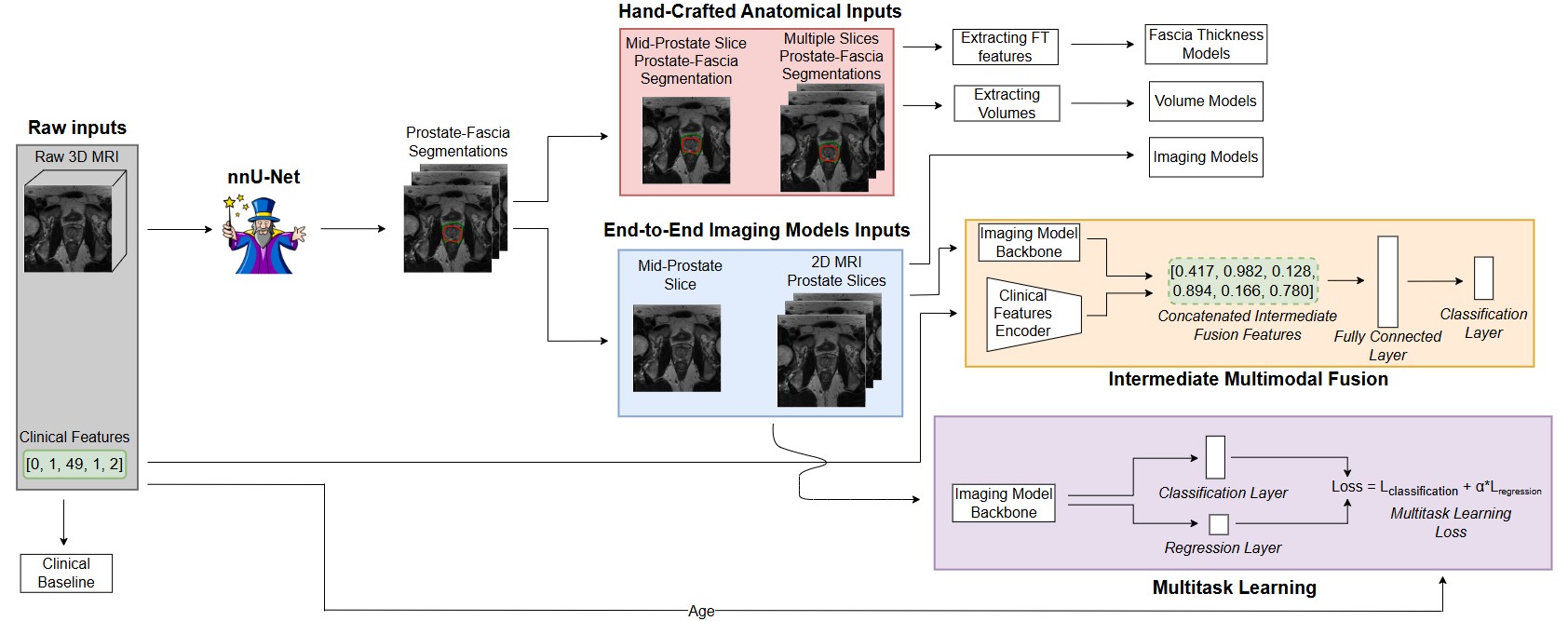}
  \caption{Schematic overview of the methodological pipeline.}
  \label{fig:schematic_pipeline}
\end{figure}

Figure \ref{fig:schematic_pipeline} presents a schematic overview. All models aim to predict ED 12 months after surgery, based on a binary target variable indicating whether the patient developed postoperative ED.

\subsubsection{Clinical Baseline}
For the clinical baseline, we use clinical features available prior to surgery: age, height, weight, smoking status and frequency, alcohol use and consumption, medication usage, comorbidities, and preoperative erectile function score. Categorical variables were imputed using the mode, and numerical ones with the mean. A set of different classical machine learning (ML) models will be evaluated, with the best-performing model serving as the strongest clinical-only baseline for comparison.

\subsubsection{Handcrafted MRI Features}
\begin{figure}[H]
\centering
\begin{minipage}[b]{0.48\linewidth}
  \centering
  \includegraphics[width=\linewidth]{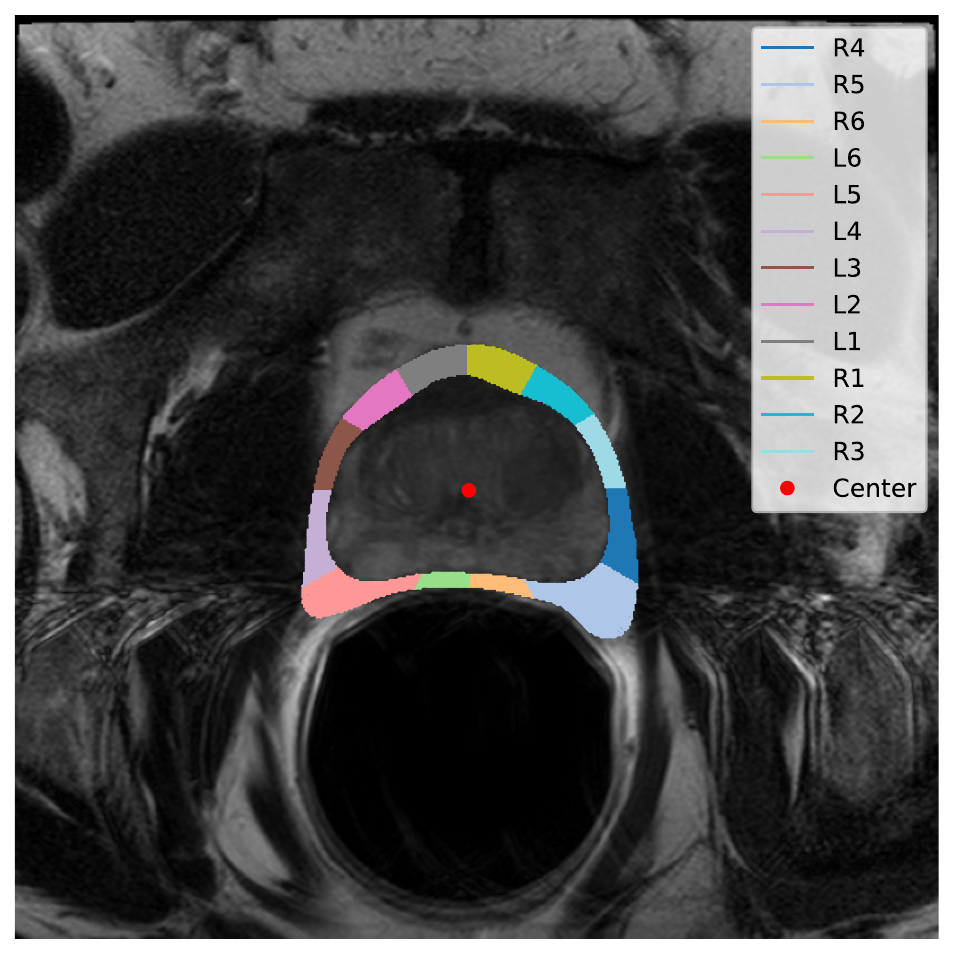}
\end{minipage}
\hfill
\begin{minipage}[b]{0.48\linewidth}
  \centering
  \includegraphics[width=\linewidth]{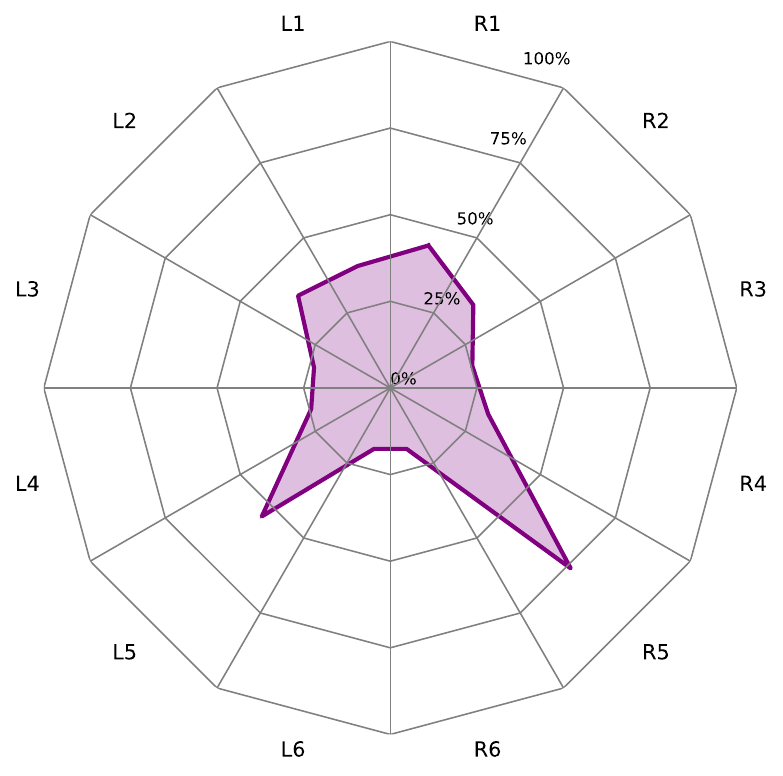}
\end{minipage}
\caption{Mid-prostate MRI slice fascia thickness analysis: (left) prostate segmented into 12 radial regions, (right) corresponding median fascia thicknesses.}
\label{fig:fascia_thick}
\end{figure}

To extract anatomical features from MRI, we focused on fascia thickness and volumetric estimates. Following Grivas et al. \cite{grivas2019value}, fascia thickness was computed on a single mid-prostate slice by dividing it into 12 radial regions of 30$^{\circ}$ each. In each region, 30 fascia thickness measurements, 1$^{\circ}$ each, were taken, and their median was used as the regional fascia thickness. Figure \ref{fig:fascia_thick} shows the radial division and corresponding fascia thickness values. We extended this analysis by applying the same procedure to 12 slices covering the prostate, yielding a multi-slice version of the fascia thickness features.

Finally, we estimated prostate and fascia volumes from the same twelve slices using pixel area and slice thickness. These three feature sets: single-slice thickness, multi-slice thickness, and volume, were each used to train classical ML models.

As expert annotations were limited, we trained nnU-Net \cite{isensee2018nnu} to segment the prostate and fascia, enabling automated feature extraction for all patients.

\subsubsection{End-to-end Imaging Models}
To assess whether DL can capture predictive imaging features beyond handcrafted anatomical measurements, we trained end-to-end models directly on preoperative MRI. Due to limited dataset size, class imbalance, and acquisition variability, such as scanner type, coil usage, and slice count, we restrict all experiments to 2D slices. These constraints make 3D modeling impractical, as it would likely lead to unstable training and poor generalization.

We evaluated several 2D input configurations: (1) a single mid-prostate slice, as in the handcrafted MRI features, (2) four mid-prostate slices, (3) eight base and mid slices, and (4) twelve consecutive prostate slices. This allowed us to assess the effect of spatial context, while accounting for expert feedback that segmentation quality declines near the apex. The multi-slice setup can also be interpreted as a structural form of data augmentation, providing spatially distinct views of the same patient to enhance model robustness and generalization. All models, ResNet-18 \cite{he2016deep}, ViT-B/16 \cite{dosovitskiy2020image}, and Hybrid ResNet-ViT \cite{yan2025hybrid}, are trained on these 2D inputs. Each slice is treated independently during training, while evaluation is performed solely on the mid-prostate slice.

\subsubsection{Multimodal Fusion}
To assess whether preoperative MRI provides complementary predictive information to known clinical features, we implemented an intermediate fusion strategy. By jointly training on both modalities and enabling their representations to interact in a shared feature space, the model can potentially learn cross-modal relationships.

We adopted the approach of Venugopalan et al. \cite{venugopalan2021multimodal}, using the best-performing imaging architecture and slice configuration to encode the MRI features. A two-layer MLP encoded the clinical features into a vector of equal dimensionality. Both embeddings are concatenated and passed through a fully connected layer, followed by a classification head. Matching dimensionality is used to promote a balanced contribution from both modalities in the fused representation.

Beyond intermediate fusion, we explore multitask learning (MTL) by training the imaging model to classify postoperative ED while simultaneously predicting patient age. As age is a strong clinical predictor of ED, this auxiliary regression task could implicitly integrate clinical knowledge into the imaging pathway. Instead of using explicit clinical inputs, the model is regularized to retain potential age-predictive features in the MRI, promoting clinically meaningful supervision and better generalization.

\section{Experiments and Results}
\subsection{Dataset Description}
For this study, a private dataset from the Netherlands Cancer Institute was used, which included patients diagnosed with prostate cancer who underwent radical prostatectomy between 2006 and 2023. From this dataset, 647 patients were selected based on the availability of a preoperative MRI, good preoperative erectile function, and a recorded postoperative score at 12 months. Erectile function was assessed using the first question of the IIEF-15 questionnaire \cite{rosen1997international}, with scores ranging from 0, no function, to 5, full function. The target variable was defined by binarizing the 12-month postoperative score, with scores of 4 or 5 categorized as 1, good function, and scores below 4 as 0, poor function.

Additionally, expert segmentations are available for 124 patients with preoperative ED. These segmentations, created by two radiologists and one urologist, all three prostate-specialized, were used to train nnU-Net.

\subsection{Preprocessing and Data Augmentations}
All MRIs and segmentations were resampled to a fixed voxel spacing of (0.273, 0.273, 2.368) mm, based on the smallest physical dimensions observed across the dataset. After data augmentation, images were centrally cropped to a standardized resolution of $512 \times 512$ pixels. Slice selection was tailored according to the specific experiments. Intensity clipping was applied between the 0.5th and 99th percentiles, followed by Z-score normalization. For 2D ViT models, images were resized to $224 \times 224$. In contrast, the CNN and Hybrid ResNet-ViT processed the original $512 \times 512$ resolution to retain spatial detail. Unlike transformers, convolutional layers can handle variable input sizes and often benefit from higher-resolution features.

To improve robustness and generalization, augmentations were applied, including random rotation, affine translation, scaling, and Gaussian blur. All augmentations were limited to the axial plane to preserve spatial consistency in the depth dimension.

\subsection{Experimental Setup}
The nnU-Net model was trained using the default settings. Test evaluation of the performance was restricted to annotated slices only, with Dice scores computed per anatomical structure (prostate and fascia). To assess anatomical variability, Dice scores were also reported per prostate region: the base, mid-prostate, and apex.

Stratified nested cross-validation was performed with 5 outer and 3 inner folds, where the inner folds are used for hyperparameter tuning using Optuna \cite{akiba2019optuna}. The same folds are used across all models. Due to missing clinical values, 139 patients (21.5\%) were excluded from clinical and multimodal models, but fold structure was retained. For each outer fold, 50 trials were evaluated over the three inner folds. Each trial was trained for up to 400 epochs with early stopping, using a patience of 50. The best configuration per fold was used to train and evaluate the model on the corresponding outer test set, resulting in five tuned models per experimental setup.

All models were trained on an NVIDIA GeForce RTX 2080 Ti using a fixed random seed for reproducibility. For DL models, the Adam optimizer was used with a decaying learning rate, weighted cross-entropy loss, where the class weights are computed from inverse label frequencies.

The multitask loss is a weighted sum of the classification loss (weighted cross-entropy) and the regression loss, with only the regression loss scaled by a hyperparameter optimized during training.

We report AUC, balanced accuracy, F1-score, and the ROC curve as evaluation metrics, with balanced accuracy serving as the early stopping criterion. The code for all experiments is publicly available on GitHub at \url{https://github.com/Trustworthy-AI-UU-NKI/Predicting-ED}.

\subsection{Results}
\subsubsection{Segmentation Results}
The performance of the nnU-Net model, across the anatomical regions and the overall mean, can be observed in Table \ref{tab:regional_segmentation}. Mid-prostate slices yielded the highest scores, while apex slices showed the lowest, likely due to poorer visibility and greater anatomical variability. 

Overall, segmentation performance was higher for the prostate, compared to the fascia, likely due to the prostate’s well-defined boundaries and higher contrast in MRI, whereas fascia is thinner, less distinct, and more susceptible to annotator disagreement. 

\begin{table}
\centering
\caption{Segmentation performance across anatomical prostate regions.}\label{tab:regional_segmentation}
\begin{tabular}{|l|l|l|}
\hline
Region & Prostate Dice & Fascia Dice \\
\hline
Base   & 0.914 $\pm$ 0.128 & 0.708 $\pm$ 0.183 \\
Mid    & 0.965 $\pm$ 0.014 & 0.738 $\pm$ 0.162 \\
Apex   & 0.864 $\pm$ 0.178 & 0.607 $\pm$ 0.303 \\
\hline
Mean   & 0.905 $\pm$ 0.040 & 0.719 $\pm$ 0.112 \\
\hline
\end{tabular}
\end{table}

\begin{figure}[H]
    \centering
    \includegraphics[width=0.32\textwidth]{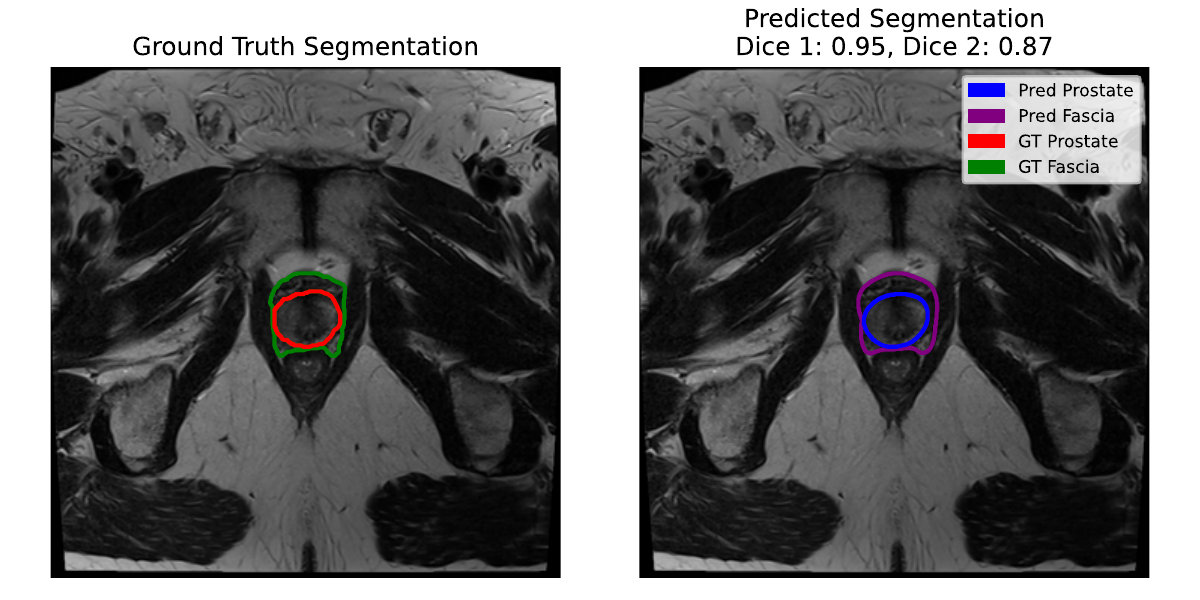}
    \includegraphics[width=0.32\textwidth]{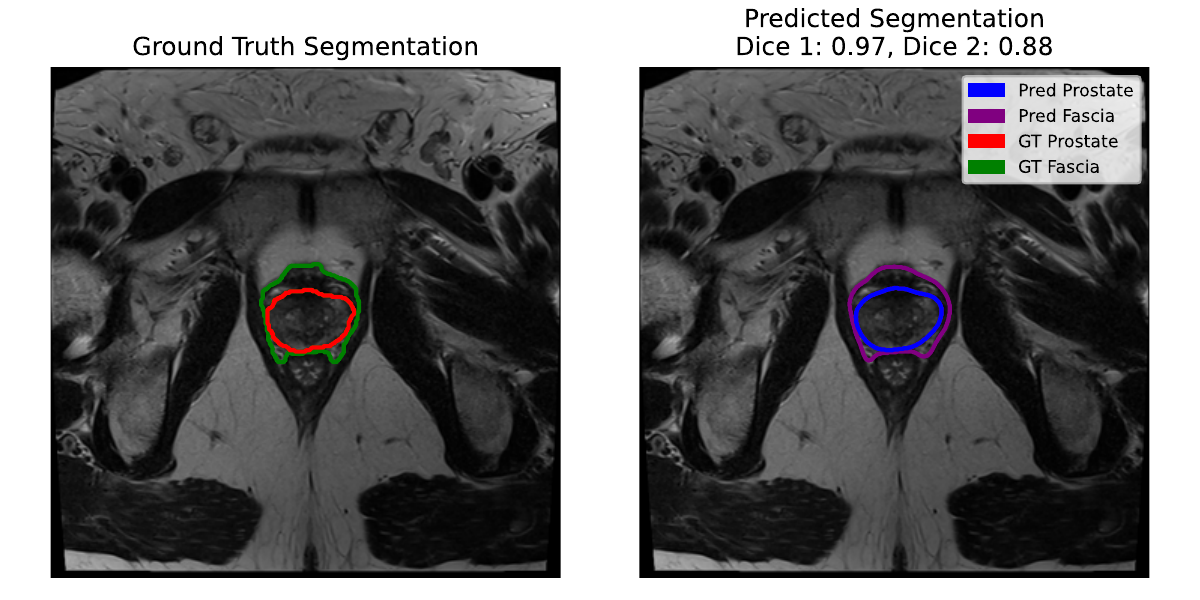}
    \includegraphics[width=0.32\textwidth]{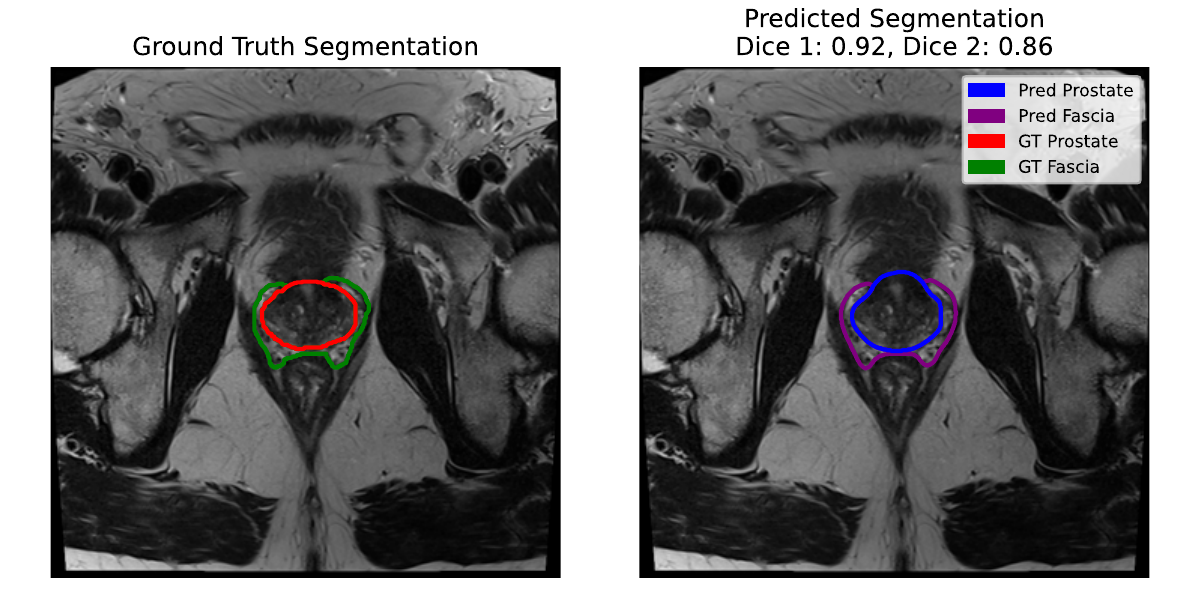}
    \caption{nnUNet segmentation results at three prostate regions. From left to right: base, mid, and apex slices.}
    \label{fig:nnunet_examples}
\end{figure}

\subsubsection{Classification Results}
\begin{figure}[H]
  \centering
  \includegraphics[width=\linewidth]{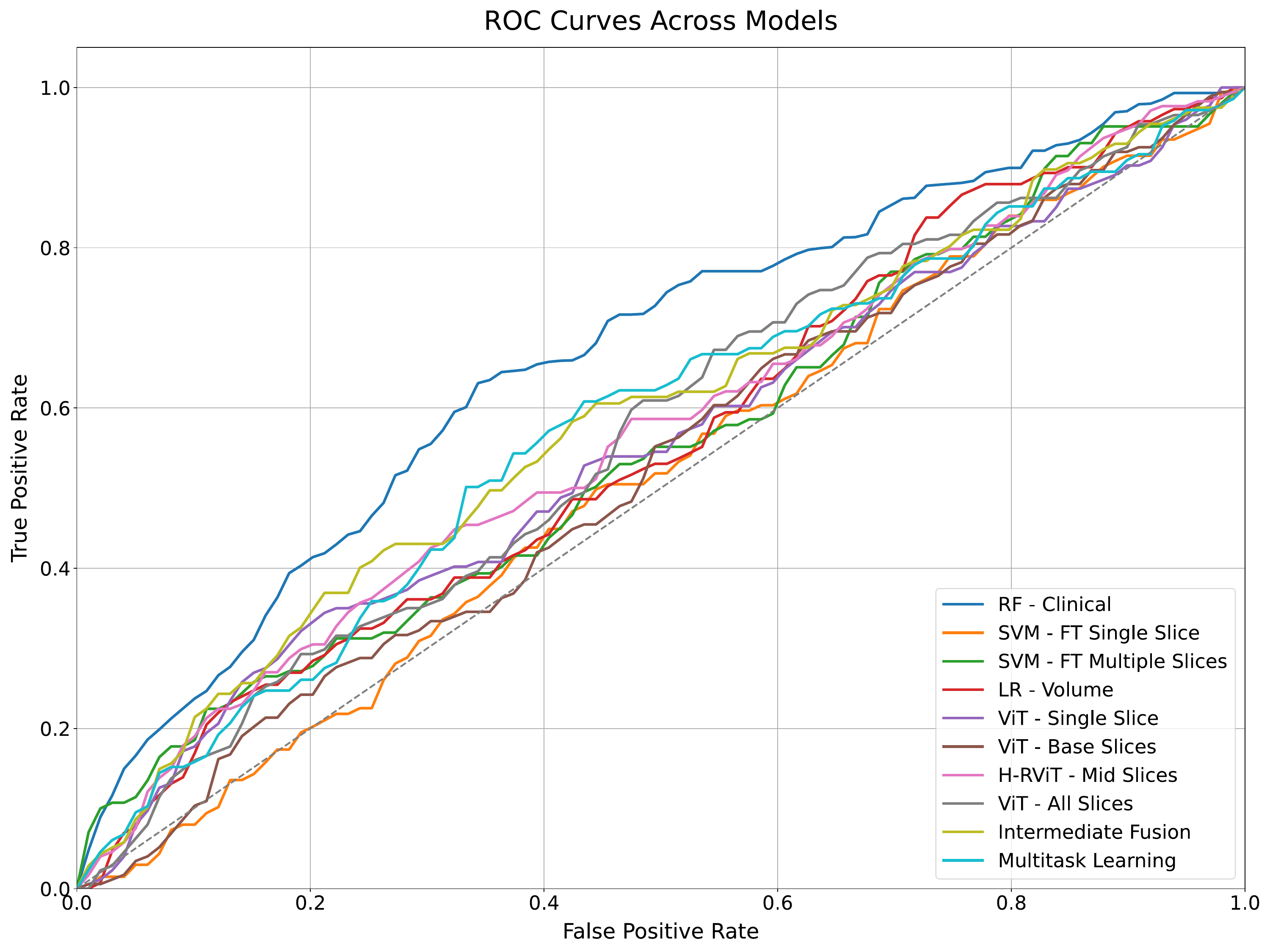}
  \caption{ROC curves of the best-performing models.}
  \label{fig:ROC_curves}
\end{figure}

We compare four modeling strategies: clinical-only baseline, handcrafted MRI features, end-to-end imaging models, and multimodal fusion. Table \ref{tab:classification_results} reports the mean AUC, balanced accuracy, and F1-score across five folds for each modeling strategy. Additionally, the ROC curves can be observed in Figure \ref{fig:ROC_curves}.

\paragraph{Clinical Baseline}
The best-performing model trained on clinical features was the Random Forest model and will serve as our benchmark for further comparison. SHAP analysis indicated that age, weight, length, medication, and preoperative erectile function score were the most important features, confirming prior work.

\paragraph{Handcrafted MRI Features}
Models trained on handcrafted fascia thickness features performed only slightly better than random. The SVM model was the best-performing model for both the single-slice and multiple-slice settings. Interestingly, the model trained on multiple-slices achieved the highest F1. Although the multiple-slice model slightly outperformed the single-slice variant numerically, the reported standard deviations suggest no statistically significant difference. The predictive value of estimated prostate and fascia volume did not seem to improve performance. Overall predictive power remained limited compared to the clinical baseline. ROC curves hovered near the diagonal, suggesting minimal discrimination.

\paragraph{End-to-end Imaging Models}
Deep learning models trained on MRI slices modestly outperformed those using handcrafted features in terms of AUC. The best performance was achieved by the pretrained Hybrid-RViT using four mid-prostate slices. Including multiple slices tended to improve robustness. Interestingly, combining base and mid slices did not improve performance compared to using only the mid-prostate slice. This may be due to increased anatomical variability or redundancy introduced by less informative regions, such as the base. In contrast, mid-prostate slices achieved the highest segmentation quality and yielded the strongest performance among imaging models, possibly suggesting they contain the most discriminative signal related to erectile function.

Nevertheless, performance remained below the clinical baseline. This underperformance may reflect a limited predictive signal in MRI, or be attributed to dataset size, acquisition variability, or anatomical heterogeneity. As shown in Figure \ref{fig:ROC_curves}, ROC curves indicate moderate improvements over random, but still underperform compared to the clinical baseline.

\paragraph{Multimodal Fusion}
The intermediate fusion model slightly improved over imaging-only models in terms of AUC and balanced accuracy, but did not surpass the clinical baseline. The results suggest that while intermediate fusion allowed for integration of multimodal features, the clinical features alone remained the most informative for predicting postoperative outcomes. The ROC curve in Figure \ref{fig:ROC_curves} confirms this, showing that intermediate fusion slightly elevates the ROC curve over imaging-only models, but does not outperform the clinical baseline. SHAP values of the concatenated imaging and clinical feature vector across the five folds indicated that the mean contribution of the clinical modality was 57.4\% of the total SHAP value ($0.00780 \pm 0.00384$), whereas the imaging modality contributed 42.6\% ($0.00579 \pm 0.00429$). These results indicate that, although both modalities contribute to the prediction, the model assigns more importance to clinical features.

The MTL model showed lower AUC and balanced accuracy but achieved a higher F1 score. Compared to imaging-only models, it performed slightly better across all metrics, except for F1, where the improvement was more pronounced. However, it still fell short of the clinical baseline in terms of AUC and balanced accuracy. These findings suggest that multitask learning may enhance the discriminative power of imaging models, particularly under data-limited conditions, by providing implicit clinical guidance.

\begin{table}[H]
\centering
\caption{Classification performance across modelling blocks (mean $\pm$ SD over five folds). Best results are shown in bold and second-best are underlined.}
\label{tab:classification_results}
\begin{tabular}{|l|l|l|l|l|}
\hline
\textbf{Method} & \textbf{Best model} & \textbf{AUC} & \textbf{Balanced Acc.} & \textbf{F1} \\
\hline
Clinical baseline & LGBM–RF & \textbf{0.663} $\pm$ 0.075 & \textbf{0.632} $\pm$ 0.061 & 0.636  $\pm$ 0.057 \\

Fascia Single & SVM & 0.513 $\pm$ 0.055 & 0.514 $\pm$ 0.037 & 0.608 $\pm$ 0.029 \\

Fascia Multiple & SVM & 0.552 $\pm$ 0.052 & 0.555 $\pm$ 0.043 & \textbf{0.687} $\pm$ 0.027 \\

Volume & LR & 0.556 $\pm$ 0.043 & 0.518 $\pm$ 0.060 & 0.501 $\pm$ 0.037 \\

DL single slice & ViT (PT) & 0.553 $\pm$ 0.080 & 0.539 $\pm$ 0.055 & 0.637 $\pm$ 0.064 \\

DL base slices & ViT (PT) & 0.526 $\pm$ 0.038 & 0.521 $\pm$ 0.039 & 0.617  $\pm$ 0.032 \\

DL mid slices & H-RViT (PT) & 0.569 $\pm$ 0.025 & 0.550 $\pm$ 0.024 & 0.642 $\pm$ 0.037 \\

DL all slices & ViT (PT) &	0.566 $\pm$ 0.049 & 0.528 $\pm$ 0.059 & 0.631 $\pm$ 0.052  \\

Intermediate fusion & \textit{N/A} &  \underline{0.586} $\pm$ 0.056  &  \underline{0.587} $\pm$ 0.029 & 0.633 $\pm$ 0.053 \\

MTL & H-RViT (PT) & 0.577 $\pm$ 0.052  & 0.561 $\pm$ 0.043 &  \underline{0.658} $\pm$ 0.023 \\
\hline
\end{tabular}
\end{table}

\subsubsection{XAI}
\begin{figure}[H]
  \centering
  \includegraphics[width=0.9\linewidth]{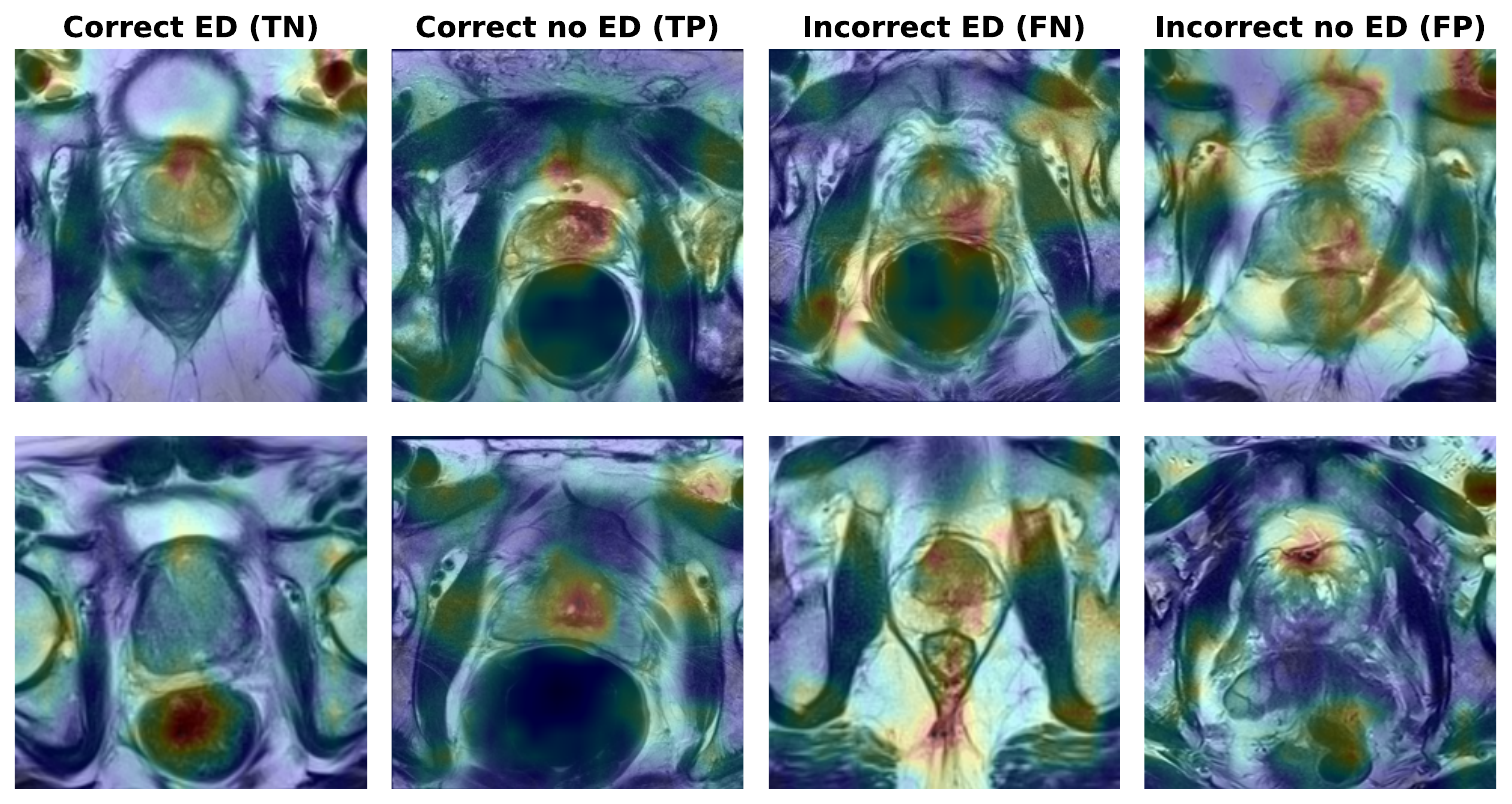}
  \caption{Attention-weighted saliency maps from the Hybrid-RViT model, highlighting focus regions in correctly and incorrectly classified MRI cases.}
  \label{fig:XAI_hybrid}
\end{figure}

To interpret what the model has learned from preoperative MRI, we visualize attention maps from the best-performing unimodal imaging model. Attention-weighted saliency maps are generated by averaging attention weights across all transformer heads and layers and multiplying them with the final convolutional feature maps. Figure \ref{fig:XAI_hybrid} shows the attention-weighted saliency maps, grouped by prediction correctness and class. Correct predictions tend to show focused, symmetric attention around the prostate, fascia, and neurovascular bundles, regions relevant to erectile function. In contrast, incorrect predictions exhibit more dispersed or anatomically implausible attention, suggesting model uncertainty or overfitting. Notably, the model occasionally highlights the endorectal coil, a non-informative structure, as well as anatomically implausible regions such as the femoral vessels, rectum, and muscular attachments, which are not directly involved in erectile function. Although the visualizations might suggest that the model has learned where to focus, the model's performance indicates limited ability to draw clinically meaningful conclusions. This highlights the need for further refinement and more diverse data to potentially improve performance.

\section{Conclusion}
This study evaluated whether preoperative MRI offers predictive value for ED 12 months after radical prostatectomy. Using a combination of handcrafted features, deep learning on MRI slices, and multimodal fusion with clinical data, we aimed to quantify the added value of imaging beyond established clinical predictors.

Our findings indicate that none of the MRI-based models outperformed the clinical baseline; however, several important insights emerged. First, deep learning models tended to perform better than handcrafted feature-based approaches and demonstrated modest predictive capability, particularly when using mid-prostate slices. Second, attention-based explainability showed that these models tend to focus on anatomically meaningful regions, such as the prostate, fascia, and neurovascular bundles, suggesting that, despite limited performance, the models start to focus on relevant spatial priors.

Multimodal fusion yielded marginal improvements over MRI-only models, particularly in F1 score, but still underperformed relative to the clinical baseline. SHAP analysis confirmed that clinical features were the dominant modality. 

These findings should be interpreted with caution due to limitations that may have affected model performance and generalizability. First, using data from a single institution may have introduced selection bias. Second, reducing erectile function to a binary outcome could have oversimplified its clinical complexity; ordinal or multiclass labels might better reflect the spectrum. Third, the dataset may have hindered stable optimization. The nested cross-validation procedure revealed large variability in the number of training epochs across folds, indicating potential instability in model convergence and contributing to inconsistent performance. Moreover, some clinically important confounders, such as diabetes, a known factor in ED outcomes as reported by Saikali et al. \cite{agochukwu2022development}, were not consistently recorded. The absence of such variables, combined with the relatively limited dataset size, likely contributed to the lower AUCs observed compared to previous studies. Future research should also explore the use of expert-labeled surgical planning regions and 3D modeling strategies to leverage the anatomical information captured in MRI more fully.

Altogether, this work highlights the current limitations of MRI-based ED prediction while offering insights and directions for future research to further establish its potential.

\subsubsection{Acknowledgements.} Research at the Netherlands Cancer Institute is supported by grants from the Dutch Cancer Society and the Dutch Ministry of Health, Welfare and Sport. The authors would like to acknowledge the Research High Performance Computing (RHPC) facility of the Netherlands Cancer Institute (NKI).
\subsubsection{Disclosure of Interests.}
The authors have no competing interests to declare that are relevant to the content of this article.

% Dataset size/kwaliteit/clinical predictors missing/Despite the absence of the diabetes variable, a smaller dataset size, and the removal of patients with preoperative ED

% Future studies could investigate the use of 3D MRI with larger, more standardized datasets to determine whether it offers a performance advantage over 2D approaches.

%TOO: grotere dataset/meer folds voor results BUITEN confidence interval


\end{document}